\documentclass[11pt]{article}
\usepackage{amsmath,epsfig,sint}
\usepackage[american]{babel}

\newcommand{\unit}{1\kern-.25em {\rm l}}

\newcommand{\be}{\begin{equation}}
\newcommand{\ee}{\end{equation}}
\newcommand{\bd}{\begin{displaymath}}
\newcommand{\ed}{\end{displaymath}}
\newcommand{\bea}{\begin{eqnarray}}
\newcommand{\eea}{\end{eqnarray}}
\newcommand{\ba}{\begin{array}}
\newcommand{\ea}{\end{array}}

\newcommand{\psibar}{\bar{\psi}}

\newcommand{\Real}{\relax{\mathsf{\Gamma\kern-.35em R}}}
\newcommand{\Int}{\relax{\mathsf{Z\kern-.40em Z}}}

\newcommand{\Nf}{N_{\rm f}}

\newcommand{\bfx}{{\bf x}}
\newcommand{\bfy}{{\bf y}}
\newcommand{\bfz}{{\bf z}}

\newcommand{\rmO}{{\rm O}}

\newcommand{\zetabar}{\bar{\zeta}}
\newcommand{\zetaprime}{\zeta\kern1pt'}
\newcommand{\zetabarprime}{\zetabar\kern1pt'}

\newcommand{\msbar}{{\rm \overline{MS\kern-0.05em}\kern0.05em}}

\def\ca{c_{\rm A}}

\def\csw{c_{\rm sw}}

\def\ctildet{\tilde{c}_{\rm t}}
\def\ctt{\ctildet}
\def\CF{C_{\rm F}}
\def\Cf{C_{\rm F}}
\def\cf{\CF}
\def\mc{m_{\rm c}}
\def\rmd{{\rm d}}
\def\fa{f_{\rm A}}
\def\fp{f_{\rm P}}

\begin{document}
  \begin{titlepage}
\begin{flushright}
DESY 05-070\\
CERN-PH-TH/2005-45\\
FTUAM-05-2\\
IFT-UAM-CSIC/05-16\\[0.5ex]
May 2005\\
\end{flushright} 

\vskip 1 cm
\begin{center}
  {\Large\bf 
A perturbative study of two four-quark operators\\[0.5ex]
in finite volume renormalization schemes\\[1ex] 
}
\end{center}
\vskip 3ex
\begin{figure}[h]
\begin{center}
\epsfig{figure=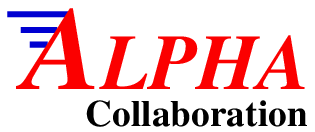,width=2.5cm} 
\end{center}
\end{figure}
\begin{center}
{\large Filippo Palombi$^{\scriptstyle a}$, Carlos Pena$^{\scriptstyle b}$ and Stefan Sint$^{\scriptstyle c}$}
\vskip 2.3ex
\begin{flushleft}
$^{\scriptstyle a}$ DESY, Theory Group, Notkestra\ss e 85, 
D-22603 Hamburg, Germany\\[1ex]
$^{\scriptstyle b}$ CERN, Theory Division, 
CH-1211 Geneva 23, Switzerland\\[1ex]
$^{\scriptstyle c}$ Departamento de F\'{\i}sica Te\'orica C-XI and
  Instituto de F\'{\i}sica Te\'orica C-XVI, Universidad 
$\hphantom{\scriptstyle cc}$Aut\'onoma de  Madrid, Cantoblanco, 
E-28049 Madrid, Spain\\[3ex]
\end{flushleft}
{\bf Abstract}
\vskip 0.7ex
\end{center}
Starting from the QCD Schr\"odinger functional (SF),
we define a family of renormalization schemes for two 
four-quark operators, which are, in the chiral limit,
protected against mixing with other operators.
With the appropriate flavour assignments these operators can be interpreted 
as part of either the $\Delta F=1$ or $\Delta F=2$ effective weak Hamiltonians.
In view of lattice QCD with Wilson-type quarks,
we focus on the parity odd components of the
operators, since these are multiplicatively renormalized 
both on the lattice and in continuum schemes.
We consider 9 different SF schemes and relate them 
to commonly used continuum schemes at one-loop order of perturbation
theory. In this way the two-loop anomalous dimensions in the SF schemes
can be inferred. As a by-product of our calculation 
we also obtain the one-loop cutoff effects in the 
step-scaling functions of the respective
renormalization constants, for both O($a$) improved and unimproved
Wilson quarks. Our results will be needed in a separate study 
of the non-perturbative scale evolution of these operators.
\vskip 3ex
\vfill
\eject
\end{titlepage}

  \section{Introduction}

In the Standard Model, four-quark operators typically arise 
as effective interaction vertices when integrating out 
the large scale physics associated with the weak interactions.
Examples are the $\Delta B=2$ operators 
\be
  O^{\Delta B=2}
 = \left(\bar b\gamma_\mu(1-\gamma_5)d\right)^2,    
\ee
which mediate $B^0$--$\bar{B}^0$ mixing or their analogues in the case of kaons.
Hence, the quantities of interest are matrix elements of
four-quark operators between hadronic states, which are
inherently non-perturbative in nature. On the other hand,  
the four-quark operators are originally obtained 
in perturbation theory and renormalized at a large scale, using
e.g.~the minimal scheme (MS) of dimensional regularization.
If lattice QCD is used for the calculation of
the hadronic matrix elements, the matching to the
perturbative renormalization schemes poses a problem,
since the scale differences involved are potentially large.
A general strategy  to solve this problem has been proposed
some time ago~\cite{Jansen:1995ck}: its starting point is the  
definition of an intermediate scheme, where the finite 
space-time volume is used to set the renormalization scale. 
Finite-size-scaling techniques then allow to step up the
energy scale recursively until the perturbative regime
is reached, where the continuum schemes can be safely matched
in perturbation theory. Previous applications 
include the running 
coupling~\cite{DellaMorte:2004bc} and 
quark mass~\cite{Sint:1998iq,Capitani:1998mq,Guagnelli:2004za},
moments of structure functions~\cite{Bucarelli:1998mu-Guagnelli:2004ga}, 
and the  static-light axial current~\cite{Kurth:2000ki,Heitger:2003xg}.

This paper is part of a project to apply this strategy to 
two particularly important four-quark operators with 
phenomenological applications to 
$B^0$--$\bar{B}^0$ or $K^0$--$\bar{K}^0$ mixing and non-leptonic 
kaon decays~\cite{Dimopoulos:2004xc-Guagnelli:2002rw}. We use the QCD Schr\"odinger functional to define
a family of finite volume renormalization schemes, and
report our results for the perturbative matching to 
more commonly used renormalization schemes.
The matching procedure is done in two steps: first 
the Schr\"odinger functional is regularized on the lattice
with Wilson-type quarks, and 
the renormalized operators are related to the standard 
lattice scheme, defined by minimal subtraction of
logarithms. Then, using results from the literature,
this lattice scheme can be related to a continuum scheme,
such as dimensional reduction (DRED) or one of the minimal subtraction
schemes ($\rm MS$) in dimensional regularization.
Since the two-loop anomalous dimensions are known in
these schemes, the one-loop matching then allows
the two-loop anomalous dimensions to be inferred in the SF schemes, too.

The paper is organized as follows: in section 2 we start with
the definition of the operators and their correlation
functions in the Schr\"odinger functional, which are
then used to formulate  the renormalization conditions.
After a short review of perturbative matching equations between 
different renormalization schemes (section 3), 
we collect the equations for the reference schemes (section 4) 
and report the results of our one-loop computation (section 5).
In view of the corresponding non-perturbative computation 
with Wilson-type quarks, we then discuss perturbative lattice artefacts 
in the step-scaling functions (section 6) and present our conclusions.

\vfill
\eject

  \section{Definitions and setup}

The four-quark operators we would like to 
renormalize are of the form
\be
  O^\pm_{\rm LL}=\frac12\left[
    (\psibar_{1}\gamma_\mu(1-\gamma_5)\psi_{2})
        (\psibar_{3}\gamma_\mu(1-\gamma_5)\psi_{4})
   \pm(\psibar_{1}\gamma_\mu(1-\gamma_5)\psi_{4})
     (\psibar_{3}\gamma_\mu(1-\gamma_5)\psi_{2})\right],
 \label{eq:LL-operators}
\ee
where the subscript ${\rm LL}$ refers to the Dirac structure
of two left-handed currents, and it is understood 
that colour indices are contracted within 
the quark bilinears in round brackets. In order to make contact
with phenomenological applications, one just needs
to assign the physical quark flavours. For instance, 
with the identifications
\be
   \psi_1=\psi_3=s,\qquad \psi_2=\psi_4=d,
\ee
the operator $O^-_{\rm LL}$ vanishes while the matrix elements
of $O^+_{\rm LL}$ appear in the $K^0$--$\bar{K}^0$ mixing amplitude. Replacing 
strange by bottom quarks, the same operator 
mediates $B^0$--$\bar{B}^0$ mixing. If instead one identifies
\be
      \psi_1=s,\qquad \psi_2=d, \qquad\psi_3=\psi_4=u,c,
\ee
one obtains the $\Delta S=1$ operators relevant to the
$\Delta I=1/2$ rule in non-leptonic kaon decays, in 
a framework where the charm quark remains an active degree of
freedom. 

The operators~(\ref{eq:LL-operators}) can be decomposed in parity-even and
-odd components:
\be
   O^\pm_{\rm LL} \equiv O^\pm_{\rm (V-A)(V-A)}=
   O^\pm_{\rm VV+AA} - O^\pm_{\rm VA+AV},
\ee
where $\rm V,A$ refer to the Dirac structure of vector and
axial vector currents, respectively.
In the following we will work with the parity-odd component of the operators,
\be
  O^\pm_{\rm VA+AV}=\frac12\left[
      (\psibar_{1}\gamma_\mu\psi_{2})
        (\psibar_{3}\gamma_\mu\gamma_5\psi_{4})
   +(\psibar_{1}\gamma_\mu\gamma_5\psi_{2})(\psibar_{3}\gamma_\mu\psi_{4})
   \pm (\psi_2 \leftrightarrow \psi_4) \right],
 \label{eq:operators}
\ee
rather than staying with the product of two left-handed 
currents (\ref{eq:LL-operators}) as induced by the Standard Model 
weak interactions.
Note that in regularizations that preserve chiral symmetry, 
parity-even and parity-odd components are related by chiral
symmetry and are thus renormalized in the same way.
However, the situation changes in lattice regularizations 
with Wilson-type quarks: while the parity-odd
operators $O^\pm_{\rm VA+AV}$ are multiplicatively 
renormalizable~\cite{Donini:1995zd}, 
their parity-even partners $O^\pm_{\rm VV+AA}$ share the lattice symmetries 
with four other four-quark operators, which leads to a
complicated operator mixing problem. Although this problem
can be solved non-perturbatively~\cite{Donini:1999sf}, 
it is a source of additional uncertainties and systematic errors.
In this paper this problem is circumvented by imposing renormalization
conditions on the parity-odd operator components. 
Besides its technical advantage
this strategy makes sense also from a practical point of
view: as has been demonstrated in~\cite{Frezzotti:2000nk}, the 
introduction of non-standard chirally twisted mass terms (``twisted mass
QCD'') redefines the physical parity symmetry so that
the parity-odd operator components can play the r\^ole of
operators with either physical parity.
\footnote{For an alternative solution using axial 
          Ward identities, see~\cite{Becirevic:2000cy}.}
As a result, the hadronic matrix elements of operators 
with even physical parity can be obtained from correlation functions
which only involve the lattice operators~(\ref{eq:operators}).
We conclude that our choice to renormalize these
parity-odd operator components is irrelevant for 
chirally symmetric regularizations, but it
is advantageous with Wilson-type quarks and does not
imply any prejudice on possible phenomenological applications.

\subsection{Renormalization conditions}

We now choose the lattice regularization with Wilson quarks,
possibly O($a$) improved by the Sheikholeslami--Wohlert term
in the action~\cite{SW}. For unexplained notation we refer the reader to
ref.~\cite{Luscher:1996sc}.
We assume that the bare operators (\ref{eq:operators})
are defined locally on the lattice, 
i.e.~all quark and antiquark fields are taken at the 
same space-time point.  
To formulate renormalization conditions 
for the renormalized operators
\be
    (O_{\rm R})^\pm_{\rm VA+AV}= Z^\pm_{\rm VA+AV} O^\pm_{\rm VA+AV},
\ee
we use the standard set-up of the 
Schr\"odinger functional~\cite{Luscher:1992an,Sint:1993un}
as described in~\cite{Luscher:1996sc}. 
We consider generic source fields made up of 
boundary quarks and antiquarks,
\begin{eqnarray}
  {\cal O}_{12}[\Gamma] &=& a^6\sum_{\rm y,z}
  \zetabar_{1}(\bfy)\Gamma\zeta_{2}({\bfz}),\\
  {\cal O}'_{12}[\Gamma] &=& a^6\sum_{\rm y,z}
  \zetabarprime_{1}(\bfy)\Gamma\zetaprime_{2}({\bfz}),
\end{eqnarray}
where $\Gamma$ is a Dirac matrix that must anticommute
with $\gamma_0$ as otherwise the source field vanishes.
This is due to the projectors $P_\pm = \frac12(1\pm\gamma_0)$,
which are implicit in the boundary quark and antiquark fields,
\be
  \zeta(\bfx)= P_- \zeta(\bfx),\qquad
   \zetabar(\bfx)= \zetabar(\bfx)P_+,
\ee
and similarly for the primed fields.
The presence of the projectors limits the possible Dirac structures 
for the quark bilinears at the time boundaries. 
For instance, it is not possible to define scalar quark bilinear 
sources, unless one introduces a non-vanishing angular momentum.
However, finite momenta typically increase the lattice artefacts, 
and lead to poor signal-to-noise ratios in numerical simulations,
so that we are not going to pursue this further.

The renormalization conditions will be imposed in the
massless theory, so that the renormalization constants and
the renormalized coupling are quark mass independent by construction.
In the absence of chirally twisted mass terms, standard parity 
is an exact symmetry of the lattice-regularized 
Schr\"odinger functional. In order to renormalize the parity-odd
operators~(\ref{eq:operators}) we thus need a total source with 
negative parity which contains at least 2 quark bilinear sources.
Because of the above mentioned problem with the
projectors at the time boundaries, we decided 
to introduce a fifth ``spectator quark'' 
and use correlation functions of the generic form:
\be
 F^\pm[\Gamma_A,\Gamma_B,\Gamma_C](x_0)
  = L^{-3}\left\langle {\cal O}_{21}[\Gamma_A] O^\pm_{\rm VA+AV}(x)
 {\cal O}_{45}[\Gamma_B]{\cal O}'_{53}[\Gamma_C] \right\rangle.
\ee
The corresponding Feynman diagram
is given in the left of fig.~1, where the quark lines correspond to 
the boundary-to-volume quark propagators $H(x),H'(x)$ 
of refs.~\cite{Luscher:1996vw,Luscher:1996jn},
and to the boundary-to-boundary propagator 
$K$ of ref.~\cite{Sint:1997jx}, which contains 
an explicit time-like link variable from Euclidean time $T-a$ to $T$.

We then consider the 5 specific cases
\begin{eqnarray}
  F^\pm_{1}(x_0) &=&  F^\pm[\gamma_5,\gamma_5,\gamma_5](x_0),\\
  F^\pm_{2}(x_0) &=& \frac16\sum_{i,j,k=1}^3 \varepsilon_{ijk} 
                     F^\pm[\gamma_i,\gamma_j,\gamma_k](x_0),\\
  F^\pm_{3}(x_0) &=& \frac13\sum_{k=1}^3 F^\pm[\gamma_5,\gamma_k,\gamma_k](x_0),\\
  F^\pm_{4}(x_0) &=& \frac13\sum_{k=1}^3 F^\pm[\gamma_k,\gamma_5,\gamma_k](x_0),\\
  F^\pm_{5}(x_0) &=& \frac13\sum_{k=1}^3 F^\pm[\gamma_k,\gamma_k,\gamma_5](x_0).
\end{eqnarray}
These correlation functions describe transitions between  
parity-even and -odd states at the Euclidean time boundaries, 
mediated by the parity-odd four-quark operators. 
In particular, parity-even scalar or axial vector states 
are produced at the lower time boundary by taking appropriate 
combinations of pseudoscalar and vector states.

In order to obtain renormalization conditions for the
four-quark operators based on these correlation functions,
we first have to take care of the source field renormalization. As
the boundary quark and antiquark fields  are all renormalized 
multiplicatively by the same renormalization 
constant~\cite{Sint:1995rb}, this can easily be achieved
by forming appropriate ratios of correlation functions.
More precisely, with the boundary-to-boundary correlators
\bea
  f_1 &=& - \dfrac{1}{2L^6}\left\langle{\cal O}_{12}[\gamma_5]
   {\cal O}'_{21}[\gamma_5] \right\rangle,\\
  k_1 &=& - \dfrac{1}{6L^6}\sum_{k=1}^3\left\langle{\cal O}_{12}[\gamma_k]
   {\cal O}'_{21}[\gamma_k] \right\rangle,
\eea
we consider the following 9 ratios of correlation functions
\bea
   h^\pm_i(x_0) &=& \dfrac{F^\pm_{i}(x_0)}{f_1^{3/2}},
   \quad i={1,\ldots,5}\\
   h^\pm_{6}(x_0) &=& \dfrac{F^\pm_{2}(x_0)}{k_1^{3/2}},\\
   h^\pm_{i+4}(x_0) &=& \dfrac{F^\pm_{i}(x_0)}{f_1^{1/2}k_1},
   \quad i={3,4,5}.  
   \label{eq:ratios_h}
\eea
In these ratios the renormalization of the boundary fields cancels
out, i.e.~the renormalized ratios are obtained by multiplying
with the appropriate four-quark operator renormalization constant.
A renormalization condition is now obtained by choosing one of the
ratios (\ref{eq:ratios_h}),
setting the renormalized quark mass to zero, and specifying
$T/L$, $x_0/L$ and the angle $\theta$ 
of the spatial boundary conditions~\cite{Luscher:1996sc}.

\begin{figure}[!t]

\begin{center}
\epsfig{figure=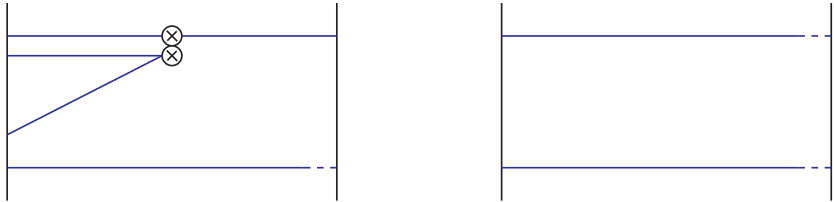, width=11 true cm}
\end{center}

\caption{The Feynman diagrams for the four-quark correlation 
functions $F^\pm_s(x_0)$ and the boundary-to-boundary correlators $f_1,k_1$ at 
tree level. Euclidean time goes from left to right. The double blob indicates 
the four-quark operator insertion and the dashed links indicate the
explicit time-like link variable in the boundary-to-boundary 
quark propagators.}
\end{figure}

All parameters are then fixed, and $L$ remains 
the only scale in the system. Then one requires
\be
    Z^\pm_{\rm VA+AV;s}(g_0,a\mu)h_s^\pm(x_0) = h_s^\pm(x_0)\vert_{g_0=0},
   \label{eq:r-condition}
\ee
where the RHS is the free field theory result.
The renormalization constant $Z^\pm_{\rm VA+AV;s}$ is thus obtained
at the renormalization scale $\mu=1/L$, and 
depends implicitly on all the parameter choices made. 
In practice, we followed refs.~\cite{Sint:1998iq,Capitani:1998mq} 
and always set $T=L$, $x_0=L/2$ and $\theta=0.5$. This 
leaves us with 9 different SF schemes labelled by $s=1,\ldots,9$
in eq.~(\ref{eq:r-condition}).
While there are good arguments for choosing $T=L$~\cite{Sint:1998iq},
there are in general no a priori criteria for a good parameter choice. 
This can only be judged a posteriori  by comparing non-perturbative and
perturbative data, or by looking at the apparent convergence
of the perturbative expansion of the anomalous dimensions.
As it will turn out, there is considerable variation 
among our 9 choices of renormalization schemes.

  \section{Anomalous operator dimensions in perturbation theory}

Operators and parameters are renormalized at the renormalization 
scale $\mu$, which in the SF schemes is identified with $1/L$.
A change of scale is then governed by the renormalization group.
Let us consider QCD with $\Nf$ mass degenerate 
quark flavours and $N$ colours, and its Euclidean correlation functions 
of gauge-invariant composite operators. Limiting ourselves to multiplicatively
renormalizable operators, any renormalized $n$-point functions
of such operators, 
\be
    G_n(x_1,\ldots,x_n) = \left\langle O_1(x_1)\cdots O_n(x_n)\right\rangle,
\ee
satisfies the Callan--Symanzik equation
\begin{equation}
  \left\{\mu\dfrac{\partial}{\partial\mu} 
  + \beta(g)\dfrac{\partial}{\partial g}
  + \tau(g) m\dfrac{\partial}{\partial m} 
 - \sum_{i=1}^n \gamma_{O_i}(g)\right\} G_n = 0.
\label{eq:CS}
\end{equation}
The renormalization group functions $\beta$, $\tau$ and $\gamma$ are 
the $\beta$-function for the coupling and the anomalous dimensions for 
the quark mass and composite operators respectively. 
In quark mass independent schemes the renormalization 
group functions only depend on the renormalized coupling and have  
perturbative expansions of the form 
\begin{eqnarray}
  \beta(g) &\buildrel{g}\rightarrow0\over\sim
            & -g^3\sum_{k=0}^\infty b_k g^{2k},\\
  \tau(g)  &\buildrel{g}\rightarrow0\over\sim
            & -g^2\sum_{k=0}^\infty d_k g^{2k},\\
  \gamma(g)  &\buildrel{g}\rightarrow0\over\sim
            & -g^2\sum_{k=0}^\infty \gamma_k g^{2k}.
\end{eqnarray}
The coefficients are scheme-dependent in general,
except for $b_0,b_1$ and $d_0,\gamma_0$.
In the normalization adopted here we have 
(see refs.~\cite{vanRitbergen:1997va,Chetyrkin:1997dh,Vermaseren:1997fq} 
for the coefficients up to $k=3$ in the $\msbar$ scheme 
and for further references):
\begin{eqnarray}
  b_0 &=& \left\{\frac{11}{3}N-\frac23\Nf\right\}(4\pi)^{-2},\\[1ex]
  d_0 &=&  \dfrac{3(N^2-1)}{N}(4\pi)^{-2},\\[1ex]
  b_1 &=& \left\{\frac{34}{3}N^2-\left(\frac{13}{3}N-N^{-1}\right)\Nf\right\}(4\pi)^{-4}.
\end{eqnarray}
Denoting the anomalous dimensions for the operators 
$O_{\rm VA+AV}^\pm$ by $\gamma^\pm$,
their leading order coefficients are given 
by~\cite{Gaillard:1974nj,Altarelli:1974ex}
\begin{equation}
   \gamma_0^\pm = \dfrac{\pm 6(N\mp 1)}{N}(4\pi)^{-2}.
\end{equation}
The universality of these coefficients
can easily be understood by changing to another quark mass independent
scheme. This amounts to finite renormalizations
of the form
\bea
  g'&=&g \sqrt{{\cal X}_{\rm g}(g)},\\[1ex]
  m'&=&m {\cal X}_{\rm m}(g),\\[1ex]
  O'_{\rm R} &=& O^{}_{\rm R}{\cal X}_{O}(g).
\eea
The $n$-point functions of the primed operators 
satisfy again a Callan--Symanzik equation of the form (\ref{eq:CS}),
and the respective renormalization group functions
are then related as follows
\begin{eqnarray}
  \beta'(g')&=& \left\{\beta(g)\dfrac{\partial g'}{\partial g}
                  \right\}_{g=g(g')},
  \label{betatransf}\\
  \tau'(g') &=& \left\{\tau(g)+\beta(g)
                  \dfrac{\partial}{\partial g}\ln{\cal X}_{\rm m}(g)
                  \right\}_{g=g(g')},\\
  \gamma'(g') &=& \left\{\gamma(g)+\beta(g)
                  \dfrac{\partial}{\partial g}\ln{\cal X}_{O}(g)
                  \right\}_{g=g(g')}.
  \label{tautransf}
\end{eqnarray}
Expanding the renormalization factors in perturbation theory,
\begin{equation}
  {\cal X}(g)\buildrel{g}\rightarrow0\over\sim
  1+\sum_{k=1}^{\infty}{\cal X}^{(k)} g^{2k}.
\end{equation}
one finds that $b_0,b_1$, and $d_0,\gamma_0$ remain indeed unchanged,
and for the next-to-leading order anomalous dimensions one arrives at
\bea
  d_1'&=& d_1 +2b_0 {\cal X}_{\rm m}^{(1)}-d_0 {\cal X}_{\rm g}^{(1)},\\
  \gamma_1' &=& \gamma_1 +2b_0 {\cal X}_{O}^{(1)}
  -\gamma_0 {\cal X}_{\rm g}^{(1)}. \label{eq:matching}
\eea
Hence, if $\gamma_1$ is known from a two-loop calculation
in some reference scheme, it can be obtained in any other scheme by 
relating the schemes at one-loop order, thereby avoiding
a direct two-loop computation. The situation 
for any multiplicatively renormalizable operator hence is the
same as with the quark mass, where the two-loop anomalous 
dimension in the SF scheme has been obtained along 
these lines~\cite{Sint:1998iq}.

  \section{Reference schemes}

The two-loop anomalous dimensions $\gamma_1^\pm$ have been computed
for a variety of schemes~\cite{Altarelli:1980fi-Buras:2000if}. 
The first computation was performed by Altarelli 
and collaborators~\cite{Altarelli:1980fi},
using dimensional reduction (DRED) \cite{Siegel:1979wq}. 
This result was later confirmed by Buras and Weisz~\cite{Buras:1989xd}, 
who used dimensional regularization with both the naive 
and the 't Hooft--Veltman  definition of $\gamma_5$ in $D$ 
dimensions~\cite{'tHooft:1972fi}.
In the DRED scheme, but with the renormalized 
coupling defined in the $\msbar$ scheme (this differs from the 
renormalized coupling used in \cite{Altarelli:1980fi}), 
the two-loop anomalous dimension takes the form
\begin{equation}
 \gamma^\pm_{1,\rm DRED} = 
   \dfrac{N\mp 1}{2N}\left\{\frac{22}{3}N^2
   -21-\frac43N\Nf\pm\left(\frac{113}{3}N
   +\frac{57}{N}-\frac{20}{3}\Nf \right)\right\}(4\pi)^{-4}.
\end{equation}
For later reference we also quote the corresponding
results in the  NDR (``dimensional regularization with naive $\gamma_5$'')
and HVDR (``dimensional regularization with 't Hooft--Veltman $\gamma_5$'')
schemes, as defined in~\cite{Buras:1989xd} 
\bea
 \gamma^\pm_{1,\rm NDR} &=& 
   \dfrac{N\mp 1}{2N}\left\{-21\pm\left(\frac{57}{N}-\frac{19}{3}N
   +\frac{4}{3}\Nf \right)\right\}(4\pi)^{-4},\\
   \gamma^\pm_{1,\rm HVDR} &=& 
   \dfrac{N\mp 1}{2N}\left\{\frac{88}{3}N^2
   -21 -\frac{16}{3}N\Nf\pm\left(\frac{157}{3}N
   +\frac{57}{N}-\frac{28}{3}\Nf \right)\right\}(4\pi)^{-4}.
\eea
In order to obtain the two-loop anomalous dimensions in the 
SF schemes, we need the one-loop relations
between the renormalized operators and coupling constants,
\bea 
     \left(O^\pm_{\rm VA+AV}\right)_{\rm SF} &=&  
     \left(O^\pm_{\rm VA+AV}\right)_{\rm DRED}     
     {\cal X}_{\rm SF,DRED}^{\pm}(\bar{g}), \\
     \bar{g}_{\rm SF}^{2}(L) &=& \bar{g}^2(\mu){\cal X}_{g}(\bar{g}).
\eea
Here we have denoted the $\msbar$ coupling by $\bar{g}$, and
we assume that the SF coupling has been defined for $N=3$ colours
as in~\cite{Luscher:1993gh,Sint:1995ch}.
There, also the one-loop coefficient for the matching of the couplings
has been determined:
\bea
   {\cal X}_{g}^{(1)} &=& 2b_0\ln(\mu L)-\dfrac{1}{4\pi}
   \left(c_{1,0}+c_{1,1}\Nf\right),\\
    c_{1,0}&=& 1.25563(4), \\
    c_{1,1}&=& 0.039863(2). 
\eea
The one-loop relation between the renormalized operators
will be established in two steps, by first converting 
to a lattice renormalization scheme ($\rm lat$),
which is obtained by minimally subtracting the logarithms~\cite{Collins}.
This yields the one-loop relation 
\be
  \left(O_{\rm VA+AV}^\pm\right)_{\rm SF} =   
  \left(O_{\rm VA+AV}^\pm\right)_{\rm lat}
  {\cal X}_{\rm SF,lat}^\pm(g_{\rm lat}),
  \label{eq:SFlat}
\ee
which will be discussed in more detail in the next section.
On the other hand, the relation between the operators in
the ${\rm lat}$-scheme and DRED has been established
in refs.~\cite{Martinelli:1983ac,Gabrielli:1990us,Frezzotti:1991pe}. 
Defining the finite renormalization factor through
\be
   (O_{\rm VA+AV}^\pm)_{\rm DRED}^{} = 
   (O_{\rm VA+AV}^\pm)_{\rm lat}{\cal X}_{\rm DRED,lat}^\pm(g_{\rm lat}),
   \label{eq:DREDlat}
\ee
the one-loop coefficient is given by
\be
    {\cal X}_{\rm DRED,lat}^{\pm (1)} = \left\{Nz_1\pm z_0+N^{-1}z_{-1}\right\}(4\pi)^{-2},
\ee
with coefficients
\bea
   z_1 &=& \frac12 \left( \Delta_{\gamma_\mu}+\Delta_{\gamma_\mu\gamma_5}
             +2\Delta_{\Sigma_1}+1\right),\\
   z_0 &=& \frac12 \left( \Delta_{\gamma_\mu}+\Delta_{\gamma_\mu\gamma_5}
             -\Delta_{1}-\Delta_{\gamma_5}-1\right),\\
   z_{-1} &=& \frac12\left(\Delta_{1}+\Delta_{\gamma_5}\right)-
             \left(\Delta_{\gamma_\mu}
              +\Delta_{\gamma_\mu\gamma_5}+\Delta_{\Sigma_1}\right).
\eea
The $\Delta$'s have been defined in \cite{Martinelli:1983ac,Gabrielli:1990us} 
and are related to the quark propagator and
vertex functions of quark bilinears. Though 
gauge parameter dependent in general, the above linear
combinations are gauge-independent, and a numerical 
evaluation yields~\cite{Martinelli:1982mw,Gabrielli:1990us,notes96},
\be
     [z_1,z_0,z_{-1}]= \begin{cases} 
           \left[-14.06090(9),\hphantom{+}5.7854(1),\hphantom{0}8.2755(2)\right] &  
             \text{for $\csw=1$},\\ 
           \left[-17.70704(7),-0.9331(1),18.6402(2)\right] &  
              \text{for $\csw=0$},
                       \end{cases} 
\ee
where to this order of perturbation theory $\csw\equiv\csw^{(0)}=0,1$ 
refers to standard and O($a$) improved Wilson
quarks, respectively. With $N=3$, we obtain
\bea
 {\cal X}_{\rm DRED,lat}^{+ (1)} &=& 
   \begin{cases} 
     -0.213020(2)& \text{for $\csw=1$},\\
     -0.302956(2)& \text{for $\csw=0$},
   \end{cases}\\
 {\cal X}_{\rm DRED,lat}^{- (1)} &=& 
   \begin{cases} 
     -0.286293(2)& \text{for $\csw=1$},\\
     -0.291138(2)& \text{for $\csw=0$}.
   \end{cases}\eea
Finally, the desired one-loop relation between the SF schemes
and the DRED scheme is obtained by combining 
eqs.~(\ref{eq:SFlat}),(\ref{eq:DREDlat}), which to one-loop order implies
\be
    {\cal X}_{\rm SF,DRED}^{\pm (1)} = {\cal X}_{\rm SF,lat}^{\pm (1)} 
    - {\cal X}_{\rm DRED,lat}^{\pm (1)}.
   \label{eq:twostep}
\ee
The  two-loop anomalous dimensions are then 
related by formula (\ref{eq:matching}), identifying the SF and
DRED schemes with the primed and unprimed schemes, respectively.

\begin{figure}[htb]

\begin{center}
\epsfig{figure=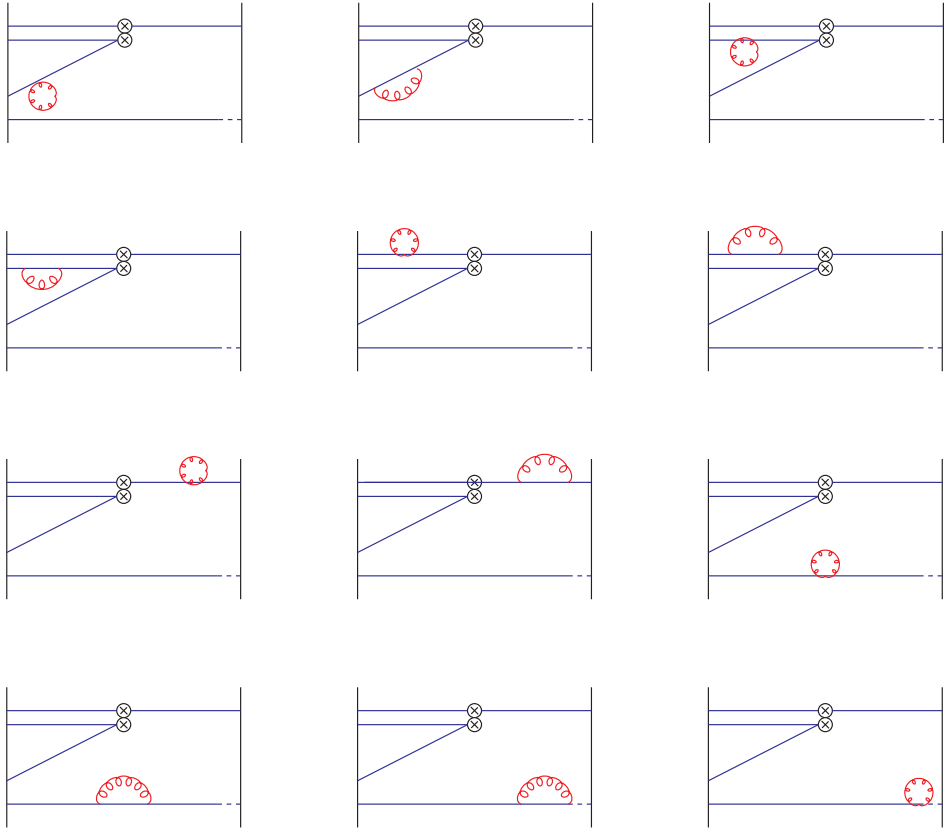, width=12.5 true cm}
\end{center}


\caption{Feynman diagrams of the self-energy type.}
\end{figure}

\begin{figure}[h]
\begin{center}
\epsfig{figure=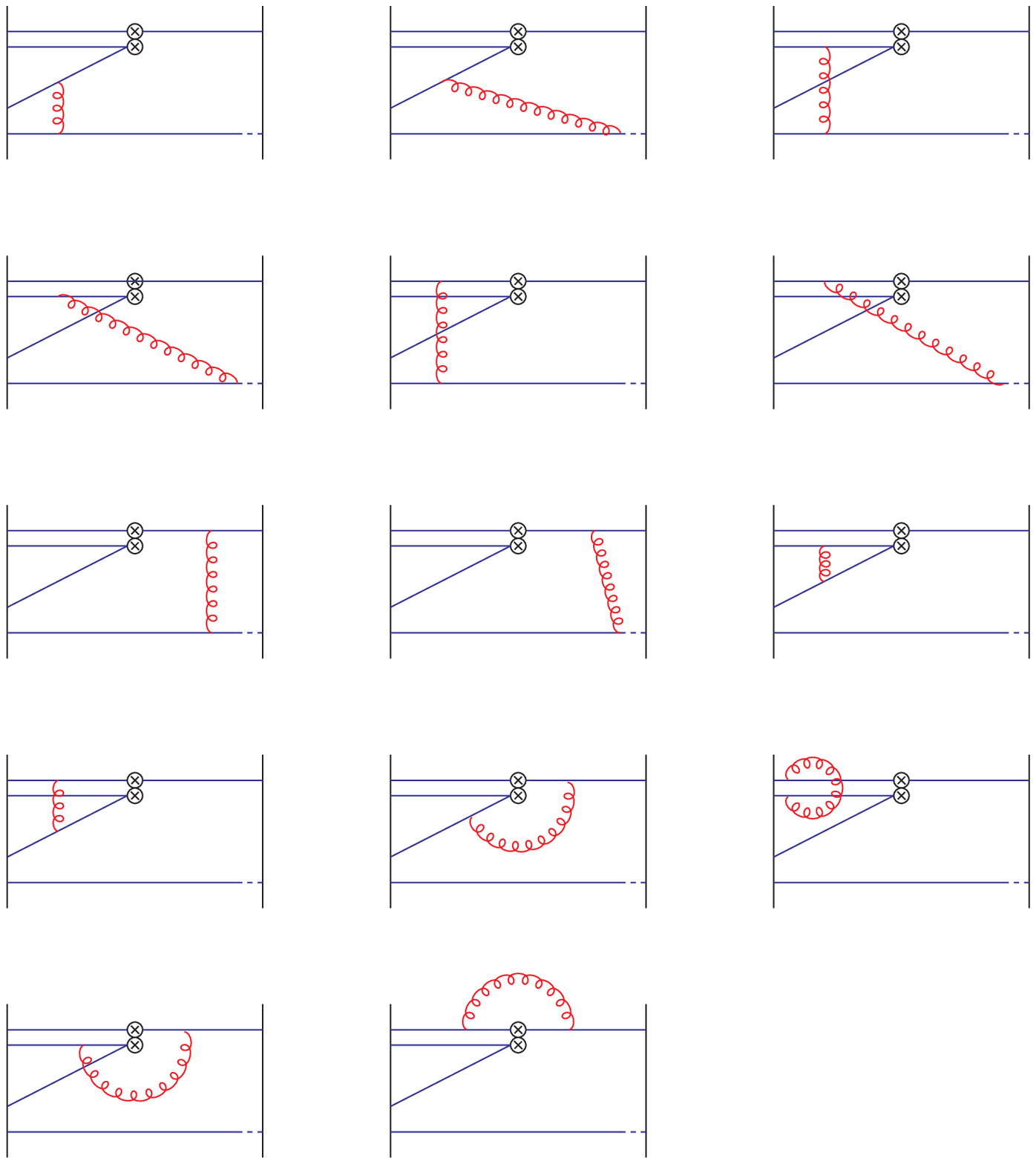, width=12.0 true cm, height=12 true cm}
\end{center}
\caption{The remaining Feynman diagrams at one-loop order,
which are not of the self-energy type.}
\end{figure}

\section{One-loop results}

The perturbative expansion of the finite volume correlation functions
is straightforward albeit a bit tedious. The technique is well-documented 
so that we refer to refs.~\cite{Luscher:1996vw,Sint:1997jx}
for details concerning the gauge-fixing procedure
and the parameter tuning necessary to take the continuum limit
while keeping the volume fixed in physical units. Here
we just describe the technical details pertaining to the 
application at hand. 
We generated double precision data for the one-loop diagrams
displayed in figs.~2 and~3, for lattice sizes ranging 
from $L/a=4$ to 32,
and we took steps of 2 in order to have a lattice 
coordinate for $x_0=L/2$. We generated data for both 
the O($a$) improved ($\csw=\csw^{(0)}=1$) and unimproved Wilson quarks 
($\csw=0$).
In the case of the O($a$) improved action, we also included the
effect of the O($a$) counterterm from the boundary proportional to
the improvement coefficient $\tilde{c}_t$~\cite{Luscher:1996sc}. 
We did not attempt to improve the four-quark operators, as
there are several O($a$) counterterms, which renders O($a$) 
improvement impractical. We note, however, that the
local operators are O($a$) improved at tree level.
Two independent sets of data  were generated by two subsets of the authors,
and perfect agreement up to rounding errors was found. We also checked
the independence of the gauge parameter, and compared with 
a numerical simulation with gauge group SU(3) at large values of 
$\beta=6/g_0^2$ (e.g.~$\beta=80$). 
As disconnected diagrams (with respect to the quark lines) 
start contributing at order $g_0^4$, the numerical values 
for these diagrams set the scale for the expected
accuracy of the comparison. 

Having passed these checks, we obtained the one-loop expressions
for the renormalization constants from the renormalization conditions
(\ref{eq:r-condition}). With the notation
\bea 
   Z^\pm_{\rm VA+AV;s}(g_0,a/L) &=& 
   1 + \sum_{n=1}^{\infty}g_0^{2n}Z^\pm_s(L/a)^{(n)},\\
   F^\pm_i(x_0) &=& \sum_{n=0}^{\infty}g_0^{2n}F^\pm_i(x_0)^{(n)}
\eea
(and analogously for $f_1,\,k_1$), we find
\bea
 Z^\pm_s(L/a)^{(1)} &=&
-\left\{\frac{F_i^\pm(L/2)^{(1)}}{F_i^\pm(L/2)^{(0)}} +
\frac{F_i^\pm(L/2)^{(1)}_b}{F_i^\pm(L/2)^{(0)}} +
m_c^{(1)}\frac{\partial}{\partial m_0}\ln F_i^\pm(L/2)^{(0)} \right\} 
\nonumber\\
&& \mbox{}+\frac{3}{2}z\left\{\frac{f_1^{(1)}}{f_1^{(0)}} +
\frac{f_{1b}^{(1)}}{f_1^{(0)}} 
+ m_c^{(1)}\frac{\partial}{\partial m_0} \ln f_1^{(0)}\right\}
\nonumber\\
&&\mbox{} +\frac{3}{2}(1-z)\left\{\frac{k_1^{(1)}}{k_1^{(0)}} +
\frac{k_{1b}^{(1)}}{k_1^{(0)}} + m_c^{(1)}
\frac{\partial}{\partial m_0} \ln k_1^{(0)}\right\}.
\eea
Here, the SF schemes $s=1$ to $s=5$ correspond to $i=1,\ldots,5$
and $z=1$, SF scheme $s=6$ translates to $i=2$ and $z=0$, and
schemes $s=7-9$ are obtained with $i=3,4,5$ and $z=1/3$.
It is assumed that the one-loop expressions are
evaluated at the bare mass $m_0=0$, i.e.~the mass counterterms
with~(see e.g.~\cite{Panagopoulos:2001fn})
\be
   am_c^{(1)} = \begin{cases} -0.20255651209\,\Cf &\text{($\csw=1$),} \\
                              -0.32571411742\,\Cf &\text{($\csw=0$),} 
                \end{cases}
                \qquad \Cf=\dfrac{N^2-1}{2N},
\label{eq:mc1}
\ee
ensure the condition of vanishing renormalized quark mass
to one-loop order. 
We have also added the contribution of the boundary counterterm 
proportional to $\ctt-1$, as indicated by the subscript $b$.
These terms only modify the O($a$) cutoff effects
and we set them to zero in the case of unimproved Wilson
quarks. 

In all cases it is expected that the one-loop coefficients $Z_s^\pm(L/a)^{(1)}$
have an asymptotic expansion in powers of the lattice spacing
of the form
\begin{equation}
  Z_s^\pm(L/a)^{(1)} 
  \simeq \sum_{\nu=0}^\infty\left(\dfrac{a}{L}\right)^\nu
  \left\{r_\nu^\pm + s_\nu^\pm\ln(L/a)\right\}.
\label{eq:r0}
\end{equation}
Here we are mainly interested in the continuum limit,
and thus the coefficients for $\nu=0$.
One expects $s^\pm_0$ to be equal to   
the universal one-loop anomalous dimension, $s_0^\pm=\gamma_0^\pm$, 
and $r_0^\pm$ is the finite part of the one-loop 
renormalization constant which determines  the one-loop matching between
the SF and the $\rm lat$ schemes,
\be
    {\cal X}^{\pm (1)}_{\rm SF,lat} = r_0^\pm. 
\ee 
We have analysed the series using standard extrapolation techniques 
\cite{Luscher:1985wf,Bode:1999sm}. 
In all cases we first checked that the coefficients $s^\pm_0$ 
of the logarithm are indeed given by the universal 
one-loop anomalous dimensions. 
Within the expected numerical precision this is indeed the case: for
instance in scheme $s=5$ and for the O($a$) improved action we get
\be
   s^+_0/\gamma_0^+ = 1.00(1),\qquad s^-_0/\gamma_0^- = 0.997(6).
\ee
Having passed this check we subtracted the logarithmic divergence using the
expected universal values $s_0^\pm=\gamma_0^\pm$.
The finite parts $r_0^\pm$ could then be obtained with 
an accuracy of  4 significant digits and are given in table~\ref{tab:r0}. 
The precision is generally better for the O($a$) improved data,
owing to the fact that O($a$) improvement of the action
together with tree-level improvement of the operators
and boundary fields implies the vanishing of the subleading
coefficients $s_1^\pm$.

\begin{table}[t]
 \centering
 \begin{tabular}{c|l l| l l}
SF scheme  &  $r_0^+(\csw=1)$  & $r_0^+(\csw=0)$ & $r_0^-(\csw=1)$  & $r_0^-(\csw=0)$ \\
 \hline
&&\\[-1.5ex]
$1$   & $-0.2444(2)$ & $-0.3346(6)$ & $-0.0635(2)$ & $-0.0683(5)$ \\ 
$2$   & $-0.2917(2)$ & $-0.3818(6)$ & $-0.1663(1)$ & $-0.1711(5)$ \\ 
$3$   & $-0.2360(2)$ & $-0.3262(5)$ & $-0.0551(1)$ & $-0.0599(5)$ \\ 
$4$   & $-0.3053(2)$ & $-0.3954(6)$ & $-0.1935(2)$ & $-0.1983(5)$ \\ 
$5$   & $-0.3004(1)$ & $-0.3905(6)$ & $-0.1754(1)$ & $-0.1801(4)$ \\ 
$6$   & $-0.3043(2)$ & $-0.3944(6)$ & $-0.1790(1)$ & $-0.1837(5)$ \\ 
$7$   & $-0.2444(2)$ & $-0.3346(6)$ & $-0.0635(2)$ & $-0.0683(6)$ \\  
$8$   & $-0.3137(1)$ & $-0.4038(6)$ & $-0.2020(1)$ & $-0.2067(6)$ \\ 
$9$   & $-0.3088(2)$ & $-0.3989(6)$ & $-0.1838(2)$ & $-0.1885(5)$ \\ 
\hline
\end{tabular}
\caption{The one-loop coefficients~(\ref{eq:r0}) 
of the renormalization constants
for the 9 SF schemes, for both improved and unimproved Wilson quarks.}
\label{tab:r0}
\end{table}

As a further check we computed the difference between the
results for improved and unimproved Wilson quarks.
These values must coincide with the ones obtained in perturbation
theory on the infinite lattice. More precisely, the
relation between the renormalized operators in
the lattice minimal subtraction schemes
\be
   (O^\pm_{\rm VA+AV})_{\rm lat({sw})} = 
   (O^\pm_{\rm VA+AV})_{\rm lat({wilson})} {\cal X}^\pm_{\rm sw,wilson}, 
\ee
can be inferred in two ways, leading, at one-loop order, to the
equations
\be
    {\cal X}^{\pm (1)}_{\rm sw,wilson} = 
    {\cal X}^{\pm (1)}_{\rm DRED,lat({wilson})} 
   -{\cal X}^{\pm (1)}_{\rm DRED,lat({sw})} 
   ={\cal X}^{\pm (1)}_{\rm SF,lat({wilson})} 
   -{\cal X}^{\pm (1)}_{\rm SF,lat({sw})}. 
\ee
Numerically, we set $N=3$ and obtain from section~4
\be
 {\cal X}^{+ (1)}_{\rm sw,wilson} = -0.089935(4), \qquad
 {\cal X}^{- (1)}_{\rm sw,wilson} = -0.004845(3).
\ee
Indeed, passing via the SF scheme reproduces these numerical values
albeit to a lesser precision.

Finally, using the coefficients in table~\ref{tab:r0} 
and combining the results according to eq.~(\ref{eq:twostep}), 
we obtain the two-loop anomalous dimensions
in the SF schemes. They are collected in table \ref{tab:gam1}, 
in units of the universal one-loop anomalous dimensions. 
We observe a large spread of numerical values,
which already suggests that not all schemes will be well-suited
for practical applications.
Concerning the  equality of the two-loop anomalous dimensions 
for the SF schemes  $1$ and $7$, there is no obvious
explanation. In particular we
do not see any reason why the two schemes should be identical
and therefore believe that the equality of the anomalous dimensions
to one-loop order is an accident. 

\begin{table}[!t]
\centering
\begin{tabular} {c | r | r }
\hline &&\\[-.5ex]
 Scheme & ${\gamma_1^+/\gamma_0^+}$  $(N=3)$ $\hphantom{01234}$& 
          ${\gamma_1^-/\gamma_0^-}$  $(N=3)$ $\hphantom{01234}$
\\[1ex]
\hline
 SF-1    & $ 0.0207(12)+ 0.00800(7)\Nf $ & $ -0.4668(6)+0.03890(4)\Nf $  \\
 SF-2    & $-0.2394(12)+ 0.02377(7)\Nf $ & $ -0.1841(3)+0.02176(2)\Nf $  \\
 SF-3    & $ 0.0669(12)+ 0.00520(7)\Nf $ & $ -0.4899(6)+0.04030(4)\Nf $  \\
 SF-4    & $-0.3142(12)+ 0.02830(7)\Nf $ & $ -0.1093(6)+0.01723(4)\Nf $  \\
 SF-5    & $-0.2873(6)+ 0.02667(4)\Nf $ & $ -0.1591(3)+0.02025(2)\Nf $  \\
 SF-6    & $-0.3087(12)+ 0.02797(7)\Nf $ & $ -0.1492(3)+0.01965(2)\Nf $  \\
 SF-7    & $ 0.0207(12)+ 0.00800(7)\Nf $ & $ -0.4668(6)+0.03890(4)\Nf $  \\
 SF-8    & $-0.3604(6)+ 0.03110(4)\Nf $ & $ -0.0860(3)+0.01581(2)\Nf $  \\
 SF-9    & $-0.3335(12)+ 0.02947(7)\Nf $ & $ -0.1360(6)+0.01885(4)\Nf $  \\
\hline 
 $\rm DRED$   & $ 0.093405 -0.0056290\Nf$ & $ 0.045911 -0.0014072\Nf$  \\
 $\rm NDR $   & $-0.011082 +0.0007036\Nf$ & $ 0.011082 +0.0007036\Nf$  \\
 $\rm HVDR$   & $ 0.221112 -0.0133688\Nf$ & $-0.035357 +0.0035181\Nf$  \\
\hline 
\end{tabular}
\caption{The two-loop anomalous dimensions in units of the
corresponding universal one-loop coefficients,
in various renormalization schemes.}
\label{tab:gam1}
\end{table}

  \begin{table}[t]
 \centering
 \begin{tabular}{r|c|c}
$L/a$  &  $a\mc^{(1)}(L/a)\vert_{\csw=1}/\cf$ & $a\mc^{(1)}(L/a)\vert_{\csw=0}/\cf$ \\[.5ex]
 \hline
&&\\[-1.5ex]
6  & $-0.20321867995$ & $-0.31794582875$ \\
8  & $-0.20265948108$ & $-0.32108637617$ \\
10 & $-0.20257791198$ & $-0.32267336579$ \\
12 & $-0.20256208759$ & $-0.32357398613$ \\
14 & $-0.20255806667$ & $-0.32412959473$ \\
16 & $-0.20255683599$ & $-0.32449503382$ \\
18 & $-0.20255642494$ & $-0.32474770548$ \\
20 & $-0.20255629201$ & $-0.32492948187$ \\
22 & $-0.20255626132$ & $-0.32506452914$ \\
24 & $-0.20255626893$ & $-0.32516755709$ \\
26 & $-0.20255628989$ & $-0.32524792419$ \\
28 & $-0.20255631414$ & $-0.32531180974$ \\
30 & $-0.20255633768$ & $-0.32536342485$ \\
32 & $-0.20255635903$ & $-0.32540571847$ \\[1ex]
\hline
$\infty$ 
   & $-0.20255651209$ & $-0.32571411742$ \\
\hline
\end{tabular}
\caption{The one-loop coefficients of the critical mass as obtained from 
the PCAC condition at finite lattice size. For the parameter choices
made here, the convergence to the values at infinite lattice size
is quadratic/cubic  in $(a/L)$, for standard/O(a) improved Wilson quarks.}
\label{tab:mc1}
\end{table}

\section{The step-scaling functions}

Beyond perturbation theory, the running of parameters
and renormalization constants is traced by computing the 
corresponding step-scaling functions. For the multiplicatively 
renormalizable operators (\ref{eq:operators}) these
are denoted by $\sigma^\pm$, and defined in the continuum limit by
\be
   \sigma_s^\pm(u)=\lim_{a\rightarrow 0}\Sigma_s^\pm(u,a/L),
   \qquad \Sigma_s^\pm(u,a/L)  =
   \left.\dfrac{Z^\pm_{\rm VA+AV;s}(g_0,a/2L)}{Z^\pm_{\rm VA+AV;s}(g_0,a/L)}
   \right\vert_{\bar{g}^2(L)=u}.
\label{eq:def_sigma}
\ee
Here again, the renormalized quark mass has been set to zero.
Similarly, the step-scaling function for the running coupling
in the SF scheme is defined by
\be
    \sigma(u)=\bar{g}^2(2L),\qquad u=\bar{g}^2(L).
\ee
The connection to the renormalization group functions $\beta,\gamma^\pm$
is then given by the two coupled equations:
\be
   \sigma^\pm(g^2) = \exp\left\{\int_g^{\sqrt{\sigma(g^2)}}\!\!\rmd g' 
    \dfrac{\gamma^\pm(g')}{\beta(g')}\right\},
\qquad
   -\ln 2 = \int_g^{\sqrt{\sigma(g^2)}}\!\!\rmd g' 
    \dfrac{1}{\beta(g')}.
\ee
This implies that the first two coefficients in the perturbative expansion,
\be
   \sigma_s^\pm(u) = 1+\sigma_{s}^{\pm(1)} u + \sigma_{s}^{\pm(2)} u^2 + \rmO(u^3),
\ee
are given in terms of $b_0$ and $\gamma_0^\pm,\gamma_1^\pm$ as,
\bea
   \sigma_s^{\pm(1)} &=& \gamma_0^\pm \ln 2,\\
   \sigma_s^{\pm(2)} &=& \gamma_1^\pm \ln 2 + 
    \left[\frac12 \left(\gamma_0^\pm\right)^2+b_0\gamma_0^\pm\right](\ln 2)^2.
\eea
\begin{table}[t]
 \centering
 \begin{tabular}{c|r|r|r|r|r|r|}
  &  $L=6a$ & $L=8a$ & $L=10a$ & $L=12a$ & $L=14a$ & $L=16a$ \\
 \hline
$\delta^+_{1}$ &
 $-0.36527$  &  $-0.23780$  &  $-0.15261$  &  $-0.10151$  &  $-0.07018$  &  $-0.05008$  \\
$\delta^+_{2}$ &
 $0.02422$  &  $-0.03333$  &  $-0.02885$  &  $-0.01913$  &  $-0.01167$  &  $-0.00653$  \\
$\delta^+_{3}$ &
 $-0.48030$  &  $-0.32171$  &  $-0.21396$  &  $-0.14771$  &  $-0.10604$  &  $-0.07863$  \\
$\delta^+_{4}$ &
 $0.21731$  &  $0.09522$  &  $0.06117$  &  $0.04699$  &  $0.03880$  &  $0.03320$  \\
$\delta^+_{5}$ &
 $0.14610$  &  $0.05147$  &  $0.03183$  &  $0.02601$  &  $0.02308$  &  $0.02099$  \\
$\delta^+_{6}$ &
 $0.19678$  &  $0.09254$  &  $0.06316$  &  $0.05017$  &  $0.04211$  &  $0.03631$  \\
$\delta^+_{7}$ &
 $-0.36527$  &  $-0.23780$  &  $-0.15261$  &  $-0.10151$  &  $-0.07018$  &  $-0.05008$  \\
$\delta^+_{8}$ &
$0.33235$  &  $0.17913$  &  $0.12252$  &  $0.09319$  &  $0.07466$  &  $0.06176$  \\
$\delta^+_{9}$ &
 $0.26114$  &  $0.13538$  &  $0.09317$  &  $0.07221$  &  $0.05894$  &  $0.04955$  \\
\hline
$\delta^-_{1}$ &
 $0.76256$  &  $0.45542$  &  $0.29780$  &  $0.20823$  &  $0.15308$  &  $0.11690$  \\
$\delta^-_{2}$ &
 $0.31555$  &  $0.20900$  &  $0.14337$  &  $0.10275$  &  $0.07664$  &  $0.05907$  \\
$\delta^-_{3}$ &
 $0.82008$  &  $0.49738$  &  $0.32847$  &  $0.23133$  &  $0.17100$  &  $0.13118$  \\
$\delta^-_{4}$ &
 $0.12247$  &  $0.08045$  &  $0.05335$  &  $0.03663$  &  $0.02617$  &  $0.01934$  \\
$\delta^-_{5}$ &
 $0.25119$  &  $0.16615$  &  $0.11336$  &  $0.08071$  &  $0.05981$  &  $0.04583$  \\
$\delta^-_{6}$ &
 $0.22927$  &  $0.14607$  &  $0.09736$  &  $0.06810$  &  $0.04975$  &  $0.03765$  \\
$\delta^-_{7}$ &
 $0.76256$  &  $0.45542$  &  $0.29780$  &  $0.20823$  &  $0.15308$  &  $0.11690$  \\
$\delta^-_{8}$ &
 $0.06495$  &  $0.03850$  &  $0.02267$  &  $0.01353$  &  $0.00824$  &  $0.00506$ \\
$\delta^-_{9}$ &
 $0.19367$  &  $0.12419$  &  $0.08269$  &  $0.05761$  &  $0.04189$  &  $0.03155$  \\
\hline
\end{tabular}
\caption{O(a) improved Wilson quarks: 
for finite lattice sizes and SF scheme $s=1,\ldots,9$, 
$\delta_{s}$ is defined as the
relative deviation of the one-loop step scaling functions 
from its continuum value, cf.~eq.(\ref{eq:delta}).}
\label{tab:artefacts_impr}
\end{table}

\begin{table}[t]
 \centering
 \begin{tabular}{c|r|r|r|r|r|r|}
  &  $L=6a$ & $L=8a$ & $L=10a$ & $L=12a$ & $L=14a$ & $L=16a$ \\
 \hline
$\delta^+_{1}$ &
 $-1.84058$  &  $-1.59016$  &  $-1.36031$  &  $-1.17808$  &  $-1.03557$  &  $-0.92259$  \\
$\delta^+_{2}$ &
 $-1.29860$  &  $-1.15706$  &  $-1.00236$  &  $-0.87333$  &  $-0.77011$  &  $-0.68725$  \\
$\delta^+_{3}$ &
 $-1.77678$  &  $-1.55304$  &  $-1.33618$  &  $-1.16118$  &  $-1.02310$  &  $-0.91301$  \\
$\delta^+_{4}$ &
 $-1.38914$  &  $-1.20924$  &  $-1.03612$  &  $-0.89692$  &  $-0.78750$  &  $-0.70059$  \\
$\delta^+_{5}$ &
 $-1.36068$  &  $-1.19302$  &  $-1.02568$  &  $-0.88964$  &  $-0.78215$  &  $-0.69649$  \\
$\delta^+_{6}$ &
 $-1.39431$  &  $-1.21275$  &  $-1.03855$  &  $-0.89868$  &  $-0.78883$  &  $-0.70162$  \\
$\delta^+_{7}$ &
 $-1.84058$  &  $-1.59016$  &  $-1.36031$  &  $-1.17808$  &  $-1.03557$  &  $-0.92259$  \\
$\delta^+_{8}$ &
 $-1.45294$  &  $-1.24637$  &  $-1.06026$  &  $-0.91382$  &  $-0.79998$  &  $-0.71017$  \\
$\delta^+_{9}$ &
$-1.42448$  &  $-1.23015$  &  $-1.04981$  &  $-0.90654$  &  $-0.79462$  &  $-0.70607$  \\
\hline
$\delta^-_{1}$ &
$1.65366$  &  $1.37471$  &  $1.16129$  &  $1.00085$  &  $0.87797$  &  $0.78150$  \\
$\delta^-_{2}$ &
 $1.14358$  &  $0.96017$  &  $0.81540$  &  $0.70456$  &  $0.61875$  &  $0.55094$  \\
$\delta^-_{3}$ &
 $1.62176$  &  $1.35614$  &  $1.14922$  &  $0.99240$  &  $0.87173$  &  $0.77670$  \\
$\delta^-_{4}$ &
 $1.23412$  &  $1.01235$  &  $0.84917$  &  $0.72814$  &  $0.63614$  &  $0.56429$  \\
$\delta^-_{5}$ &
$1.17376$  &  $0.97756$  &  $0.82666$  &  $0.71242$  &  $0.62454$  &  $0.55539$  \\
$\delta^-_{6}$ &
 $1.19143$  &  $0.98801$  &  $0.83350$  &  $0.71723$  &  $0.62811$  &  $0.55813$  \\
$\delta^-_{7}$ &
$1.65366$  &  $1.37471$  &  $1.16129$  &  $1.00085$  &  $0.87797$  &  $0.78150$  \\
$\delta^-_{8}$ &
 $1.26602$  &  $1.03091$  &  $0.86124$  &  $0.73659$  &  $0.64238$  &  $0.56908$  \\
$\delta^-_{9}$ &
$1.20566$  &  $0.99612$  &  $0.83873$  &  $0.72087$  &  $0.63078$  &  $0.56018$  \\
\hline
\end{tabular}
\caption{The same as table \ref{tab:artefacts_impr},
but for unimproved Wilson quarks.
The cutoff effects are very large, mainly due to the 
lattice artefacts in the definition of $\mc$ on finite lattices 
(cf.~fig.~4).} 
\label{tab:artefacts_wils}
\end{table}

On the lattice the computation of 
the step scaling functions $\sigma_s^\pm(u)$ requires a careful
extrapolation of lattice approximants $\Sigma_s^\pm(u,a/L)$, obtained
for different lattice sizes $L/a$ at fixed values $u$.
The limit is expected to be reached at a rate proportional to $a/L$,
but in practice higher order effects may still be important
for the accessible lattice sizes.
In perturbation theory we can now address this question
by studying the continuum approach 
of the perturbative coefficients,
\be
   \Sigma_s^\pm(u,a/L) = 1+\Sigma_s^\pm(a/L)^{(1)}\, u 
   +\Sigma_s^\pm(a/L)^{(2)}\, u^2 
   + \rmO(u^3).
\ee
Our one-loop calculation allows the study of 
\be
    \Sigma_s^\pm(a/L)^{(1)} = Z_s^\pm(2L/a)^{(1)}-Z_s^\pm(L/a)^{(1)}.
\ee
This is in principle straightforward, but there is a 
subtlety related to the determination of the zero mass point.
With Wilson-type quarks, the chiral limit can only be defined 
up to cutoff effects. Perturbation theory is special in this respect, 
as $g_0$ and $a$ can be varied independently order by order in the expansion
so that a critical mass parameter can be unambiguously defined 
at each order of perturbation theory (cf.~(\ref{eq:mc1}) for 
the one-loop results).  However, for the perturbative evaluation 
of cutoff effects to be useful in the analysis of the corresponding 
non-perturbative simulation data, we would like to mimick
exactly the procedure used there. In particular we take over 
the definition of the zero mass point: on a lattice of size $L/a$,
with $T=L$ and $\theta=0$, one determines the bare mass parameter
$m_0$ for which the PCAC mass
\be
    m(x_0) = \dfrac{\fa(x_0+a)-\fa(x_0-a)}{4a\fp(x_0)}
   \label{eq:PCAC}
\ee
vanishes at the midpoint $x_0=L/2$~\cite{Capitani:1998mq}. 
In perturbation theory, the condition
$m(L/2)=0$ then leads to a perturbative series for the
critical mass including cutoff effects,
\be
    m_c(L/a)= \sum_{n=0}^\infty m_c^{(n)}(L/a)g_0^{2n},
\ee
with the low order results
\bea
   m_c^{(0)}(L/a) &=& 0,\\
   m_c^{(1)}(L/a) &=& -\left.\dfrac1{4a}\left(\fa^{(1)}(x_0+a)
                      -\fa^{(1)}(x_0-a)\right)\right\vert_{m_0=0}.
\eea
We remark that, in the O($a$) improved framework ($\csw=1$), 
the SF correlation function $\fa(x_0)$ in eq.~(\ref{eq:PCAC})
is supposed to include also the O($a$) counterterms 
proportional to $\ca$ and $\ctt$. However, at one-loop order and 
with the chosen parameters, these O($a$) counterterms vanish identically.
The limiting values of $m_c^{(1)}(L/a)$ for infinite
lattice size $L/a$ are the usual one-loop coefficients~(\ref{eq:mc1}),
which are reached with a rate proportional to $(a/L)^2$ 
and $(a/L)^3$ for Wilson and O(a) improved Wilson quarks, respectively.
For future reference we have collected the values of 
$am_c^{(1)}(L/a)$ for lattice sizes up to $L/a=32$ in table~\ref{tab:mc1}.

Having determined the critical quark mass, the 
cutoff effects in the step-scaling functions can be evaluated
in a straightforward way.
Defining the relative deviation of the one-loop coefficients
\be
   \delta_s^\pm(a/L) = \Sigma_s^\pm(a/L)^{(1)}/\sigma_s^{\pm(1)} -1,
  \label{eq:delta}
\ee
the results for the various SF schemes and both improved and unimproved
Wilson quarks are given in tables~4 and~5.
As can be seen there, cutoff effects in the one-loop coefficient with
O($a$) improved Wilson quarks are typically around the 30-50 percent 
level at $L/a=6$, and decrease to a few percent level at $L/a=16$.
With unimproved Wilson quarks, however, the cutoff effects 
are found to be much larger, 
typically going from 150 down to 60-80 percent at the 
largest lattice size. We found that most of this dramatic effect
is indeed due to the usage of $\mc(L/a)$ rather than $\mc(\infty)$.
This is illustrated in fig.~4, where the corresponding values
for $\delta^+_1$ and $\delta^-_2$ are plotted both for improved
and unimproved Wilson quarks.
\begin{figure}[h]
\hskip -0.5cm
\epsfig{figure=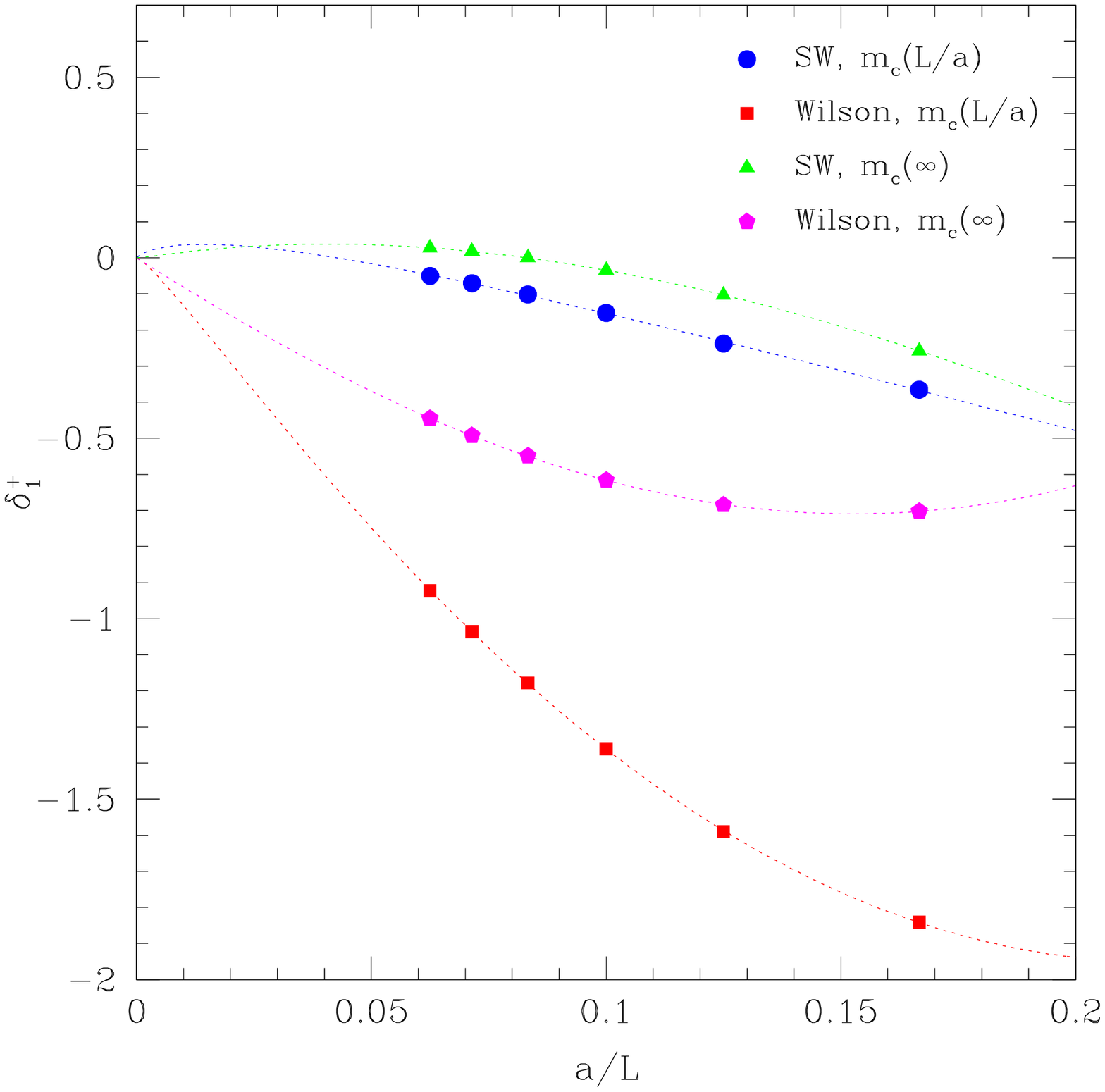, width=8 true cm}
\hskip -0.2cm
\epsfig{figure=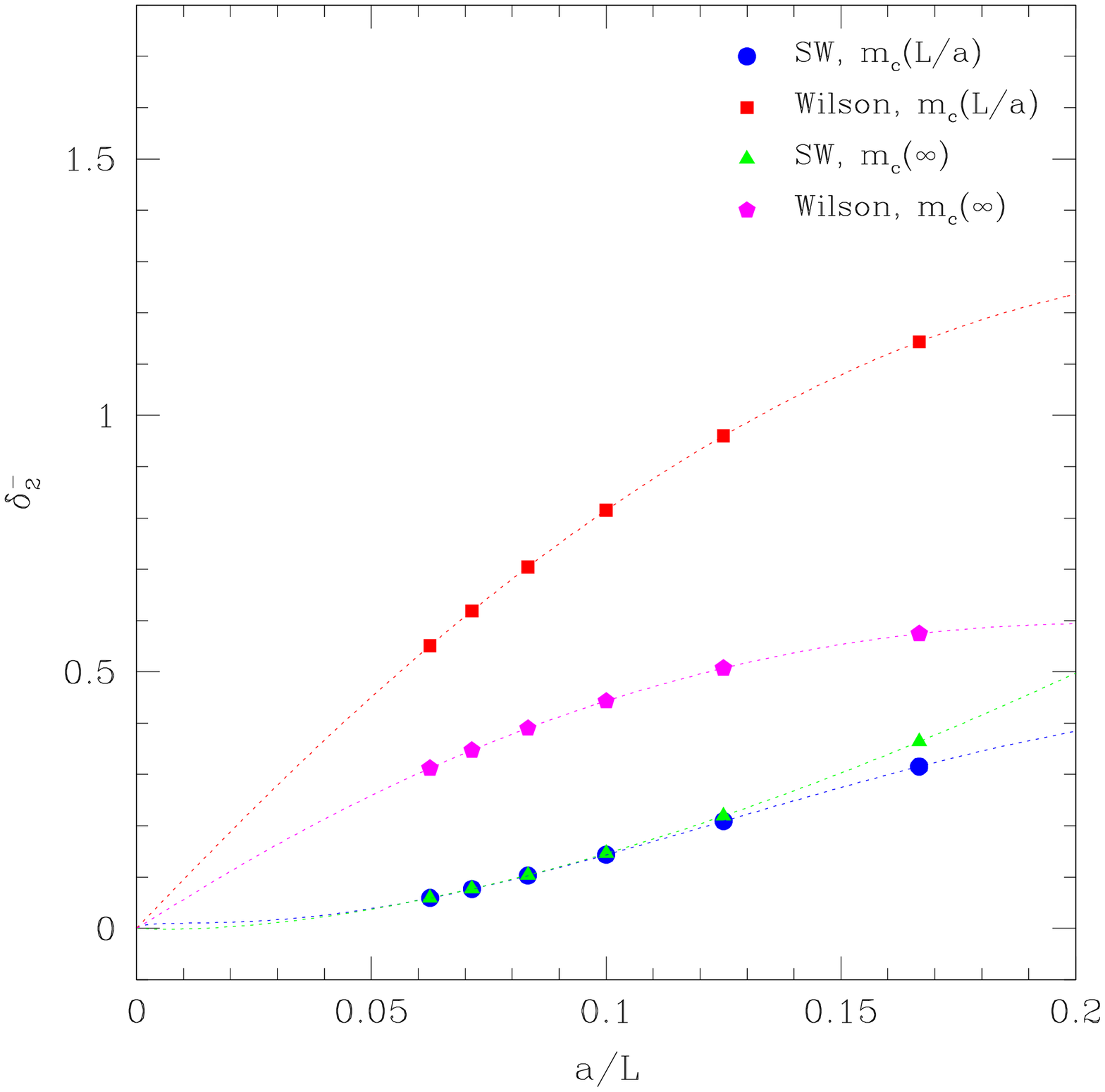, width=8 true cm}
\caption{One-loop lattice artefacts $\delta^+_1$  and $\delta^-_2$
in the step-scaling function.
The data points are obtained with the Wilson and SW actions, each
for two definitions of the critical mass. The dashed lines 
are obtained as 4-parameter fits to the expected asymptotic behaviour
and are displayed to guide the eye.}
\end{figure}

We conclude that cutoff effects in the step-scaling functions
can be quite large, and the expected asymptotic dominance
of linear lattice artefacts is not yet observed for our data.
However, we also note that cutoff effects with the O($a$) improved 
action are significantly smaller, a fact that is also reflected in the
non-perturbative data~\cite{NP}. Moreover, for the
available lattice sizes O($a^2$) effects seem to dominate.
We take this as an indication that 
O($a$) improvement of the four-quark operators
may be numerically unimportant, at least for the step-scaling functions
considered here.

  \section{Conclusions}

We have introduced a family of finite volume renormalization schemes 
for the two multiplicatively renormalizable
four-quark operators in eq.~(\ref{eq:operators}). The schemes are
based on the Schr\"odinger functional and are defined
independently of a particular regularization. By matching,
at one-loop order of perturbation theory,
to commonly used continuum schemes (NDR, HVDR, DRED),
we could infer the 2-loop anomalous operator dimensions in these SF schemes.
These results are being used in a corresponding non-perturbative
study~\cite{NP} which completely solves the non-perturbative renormalization
problem for these operators in quenched QCD.
Based on this work, preliminary results for the kaon bag parameter 
$B_K$ have been presented in~\cite{Dimopoulos:2004xc}.
Besides the two multiplicatively renormalizable operators studied here, 
the complete basis of parity-odd four-quark operators 
contains eight further operators  which form four pairs that 
mix under renormalization~\cite{Donini:1999sf}. The study of these 
mixing problems both in perturbation theory and beyond is left
for future work.

  \subsection*{Acknowledgements}
We thank A. Vladikas for many useful discussions and critical
comments on a draft of this paper. S.S and F.P.~acknowledge partial support
by the cooperation CICYT-INFN 2004. C.P. thanks both the CERN Theory
Division and the University of Rome ``Tor Vergata'' for their hospitality.

\end{document}